**Chapter 17**

# Modelling and Simulation of Fog and Edge Computing Environments using iFogSim Toolkit

***Redowan Mahmud, Rajkumar Buyya***

## 17.1 Introduction

Relying on rapid advancement of hardware and communication technology, Internet of Things (IoT) is consistently promoting every sphere of cyber-physical environments. Consequently, different IoT-enabled systems such as smart healthcare, smart city, smart home, smart factory, smart transport and smart agriculture are getting significant attention across the world. Cloud computing is considered as the base stone for offering infrastructure, platform and software services to develop IoT enabled systems [1]. However, Cloud datacenters reside at a multi-hop distance from the IoT data sources that increases latency in data propagation. This issue also adversely impacts the service delivery time of IoT enabled systems and for real time use cases such as monitoring health of critical patients, emergency fire and traffic management, it is quite unacceptable. In addition, IoT devices are geographically distributed and can generate a huge amount of data in per unit time. If every single IoT-data is sent to Cloud for processing, the global Internet will be overloaded. To overcome these challenges, involvement of Edge computational resources to serve IoT-enabled systems can be a potential solution [2].

Fog computing, interchangeably defined as Edge computing, is a very recent inclusion in the domain of computing paradigms that targets offering Cloud-like services at the edge network to assist large number of IoT devices. In Fog computing, heterogeneous devices such as Cisco IOx networking equipment, micro-datacenter, Nano-server, smart phone, personal computer and Cloudlets, commonly known as Fog node, create a wide distribution of services to process IoT-data closer to the source. Hence, Fog computing plays a significant role in minimizing the service delivery latency of different IoT-enabled systems and relaxing the network from dealing a huge amount of data-load [3]. Compared to Cloud datacenters, Fog nodes are not resource enriched. Therefore, most often, Fog and Cloud computing paradigm work in integrated manner (Figure 17.1) to tackle both resource and Quality of Service (QoS) requirements of large scale IoT-enabled systems [4].

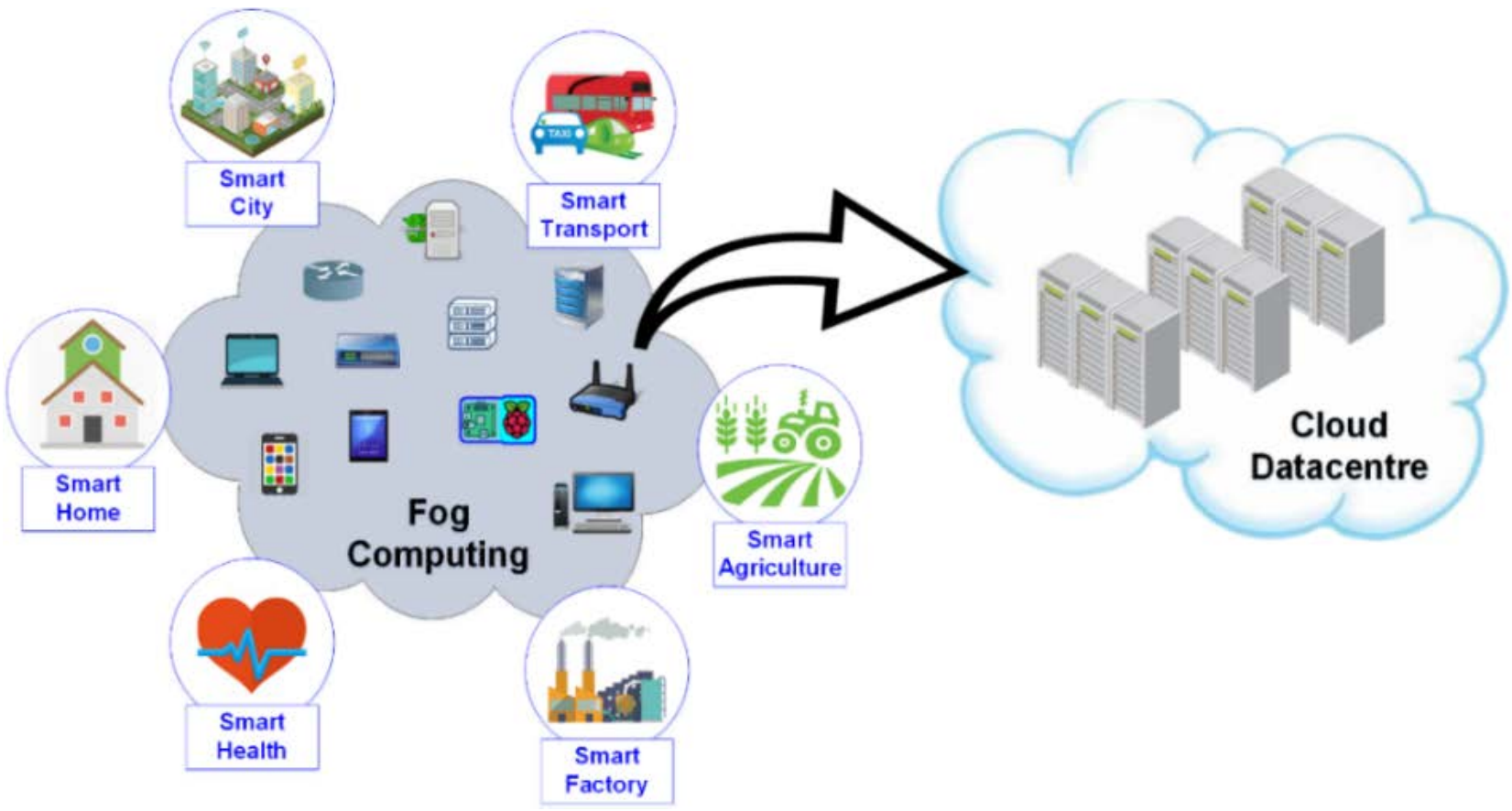

**Figure 17.1: Interactions among IoT-enabled systems, Fog and Cloud computing**

Resource management in Fog computing is very complicated as it engages significant number of diverse and resource constraint Fog nodes to meet computational demand of IoT-enabled systems in distributed manner. Its integration with Cloud triggers further difficulties in combined resource management. Different sensing frequency of IoT devices, distributed application structure and their coordination also influence resource management in Fog computing environment [5]. For advancement of Fog and its resource management, the necessity of extensive research in beyond question.

In order to develop and evaluate different ideas and resource management policies, empirical analysis on Fog environment is the key. Since Fog computing environment incorporates IoT devices, Fog nodes and Cloud datacenters along with huge amount of IoT-data and distributed applications, real-world implementation of Fog environment for research will be very costly. Moreover, modification of any entity in real-world Fog environment will be tedious. In this circumstance, simulation of Fog computing environment can be very helpful. Simulation toolkits not only provide frameworks to design customized experiment environment but also assist in repeatable evaluation. There exists a certain number of simulators such as Edgecloudsim [6], SimpleIoTSimulator [7] and iFogSim [8] for modelling Fog computing environment and running experiments. In this chapter we focus on delivering a tutorial on iFogSim. iFogSim is currently getting remarkable attention from Fog computing researchers and we believe this chapter will offer them a simplified way to apply iFogSim in their research works.

In later sections of the chapter, we briefly discuss the iFogSim simulator and its basic components. We revisit the way of installing iFogSim and provide a guideline to model Fog environment. Some Fog scenarios and their corresponding user extensions are also included in this chapter. Finally, we conclude the chapter with simulation of a simple application placement policy and a case study.

## 17.2 iFogSim Simulator and its Components

iFogSim simulation toolkit is developed upon the fundamental framework of CloudSim [9]. CloudSim is one the wildly adopted simulators to model Cloud computing environments [10] [11]. Extending the abstraction of basic CloudSim classes, iFogSim offers scopes to simulate customized Fog computing environment with large number of Fog nodes and IoT devices (e.g. sensors, actuators). However, in iFogSim the classes are annotated in such a way that users, having no prior knowledge of CloudSim, can easily define the infrastructure, service placement and resource allocation policies for Fog computing. iFogSim applies *Sense-Process-Actuate* and distributed dataflow model while simulating any application scenario in Fog computing environment. It facilitates evaluation of end to end latency, network congestion, power usage, operational expenses and QoS satisfaction. In a significant amount of research works, iFogSim has already been used for simulating resource [12], mobility [13], latency [14], Quality of Experience (QoE) [15], energy [16], security [17] and QoS-aware [18] management of Fog computing environment. iFogSim is composed of 3 basic components;

- ***Physical Components***

Physical components include Fog devices (Fog nodes). The Fog devices are orchestrated in hierarchical order. The lower level Fog devices are directly connected with associated sensors and actuators. Fog devices act like the datacentres in Cloud computing paradigm by offering memory, network and computational resources. Each Fog device is created with specific instruction processing rate and power consumption attributes (busy and idle power) that reflects its capability and energy efficiency.

The sensors in iFogSim generates tuples that can be referred as tasks in Cloud computing. The creation of tuples (tasks) is event driven and the interval between generating two tuples is set following deterministic distribution while creating the sensors.

- ***Logical Components***

Application modules (*AppModules*) and Application edges (*AppEdges*) are the logical components of iFogSim. In iFogSim applications are considered as a collection of inter-dependent AppModules that consequently promotes the concept of distributed application. The dependency between two modules are defined by the features of AppEdges. In the domain of Cloud computing, the AppModules can be mapped with Virtual Machines (VMs) and the AppEdges are the logical dataflow between two VMs. In iFogSim, each AppModule (VM) deals with a particular type of tuples (tasks) coming from the predecessor AppModule (VM) of the dataflow. The tuple forwarding between two AppModules can be periodic and upon reception of a tuple of a particular type, whether a module will trigger another tuple (different type) to next module is determined by fractional selectivity model.

- ***Management Components.***

Management component of iFogSim is consist of Controller and Module Mapping objects. The Module Mapping object according to the requirements of the AppModules, identifes available resources in the Fog devices and place them within it. By default, iFogSim support hierarchical placement of the modules. If a Fog device is unable to meet the requirements of a module, the module is sent to upper level Fog device. The Controller object launches the AppModules on their assigned Fog devices following the placement information provided by Module Mapping object and periodically manages the resources of Fog devices. When the simulation is terminated, the Controller object gather results of cost, network usage and energy consumption during the simulation period from the Fog devices. The interaction between iFogSim components are represented in Figure 17.2.

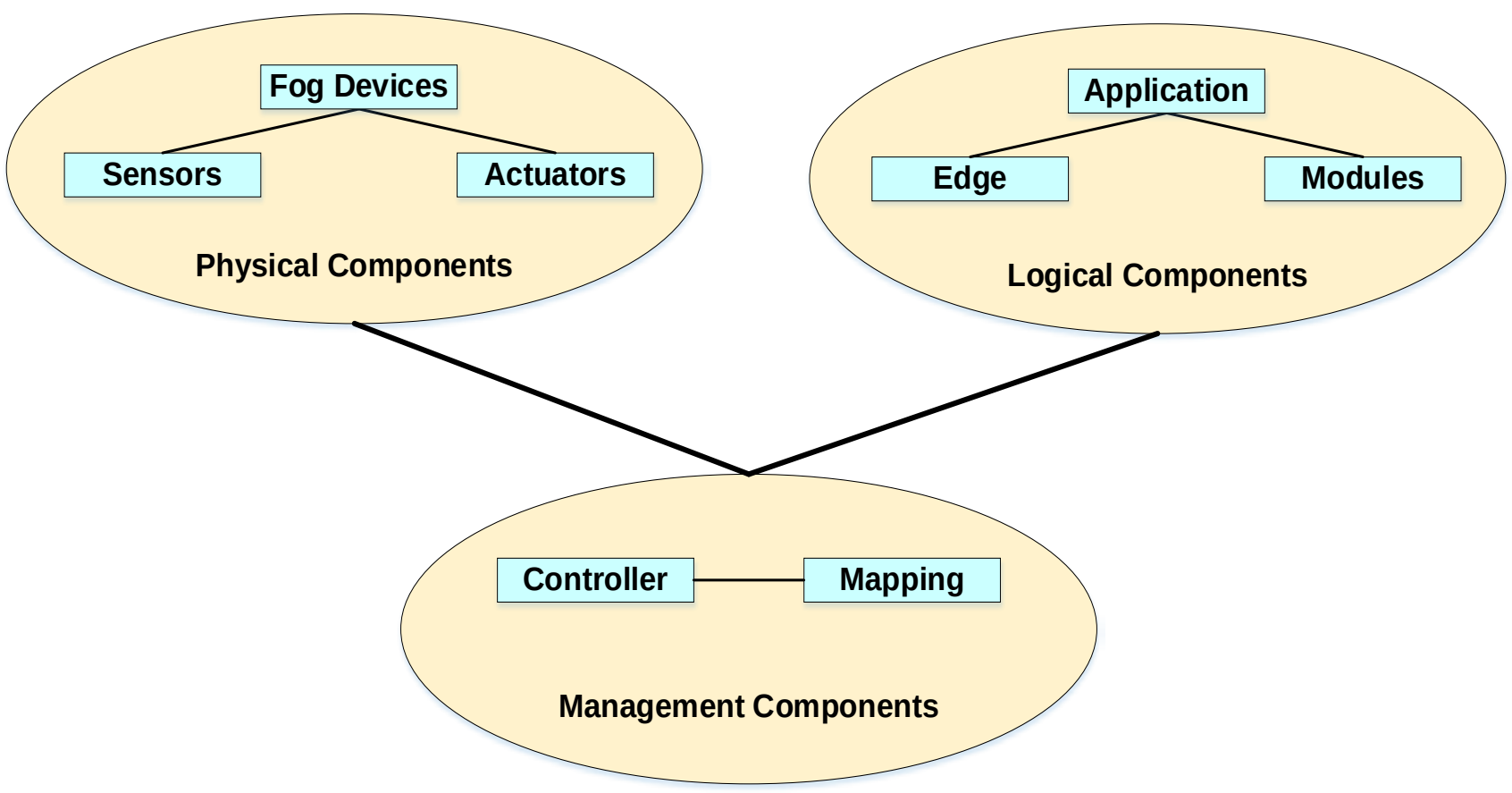

**Figure 17.2: High-level view of interactions among iFogSim components**

## 17.3 Installation of iFogSim

iFogSim is an open source Java based simulator developed by Cloud Computing and Distributed Systems (CLOUDS) Laboratory at University of Melbourne. The download link for iFogSim source code is given in their website. A very simple way to install iFogSim is described below;

1. Download iFogSim source zip file from https://github.com/Cloudslab/iFogSim or http://cloudbus.org/cloudsim/
2. Extract the zip file named *iFogSim-master*
3. Install Java Standard Edition Development Kit (jdk) / Runtime Environment (jre) 1.7 or more and Eclipse Juno / latest releases in personal computer.
4. Define workspace for Eclipse.
5. Create a folder in workspace.
6. Copy and Paste all contents from *iFogSim-master* to the newly created folder.
7. Open Eclipse application wizard and create a new Java Project with the same name of newly created folder.
8. From *src* (source) of the project, open org.fog.test.perfeval package and run any of the example simulation codes.

## 17.4 Building Simulation with iFogSim

In this section, high-level steps to model and simulate Fog computing environment in iFogSim is explored.

1. At first the physical components require to be created with specific configuration. The configuration parameters include ram, processing capability in million instructions per second (MIPS), cost per million instruction processing, uplink and down-link bandwidth, busy and idle power along with their hierarchical level. While creating lower level Fog devices, the associate IoT devices (sensors and actuators) need to be created. Particular value in *transmitDistribution* object that is set in creating a IoT sensor refers to its sensing interval. In addition, the creation of sensor and actuators require the reference of application id and broker id.

2. Next, the logical component such as AppModule, AppEdge and AppLoop are required to be created. While creating the AppModules, their configurations are provided and the AppEdge objects include information regarding tuple's type, their direction, CPU and networking length along with the reference of source and destination module. In background, different types of tuples are created based on the given specification on AppEdge objects.
3. Finally, Management Components (Module Mapping) are initiated to define different scheduling and AppModule placement policies. Users can consider total energy consumption, service latency, network usage, operational cost and device heterogeneity while assigning AppModules to Fog devices and can extend the abstraction of Module Mapping class accordingly. Based on the information of AppEdges, the requirements of an AppModule need to be aligned with the specification of corresponding tuple type and satisfied by the available Fog resources. Once the mapping of AppModules and Fog devices are conducted, the information of physical and logical components are forwarded to Controller object. The Controller object later submit the whole system to CloudSim engine for simulation.

# 17.5 Example Scenarios

To start with iFogSim, it is recommended to follow the built-in example codes such as VRGameFog and DCNSFog . Here, we discuss several Fog environment scenarios that can be simulated through iFogSim.

### *17.5.1 Create Fog nodes with heterogeneous configurations.*

The *FogDevice* class of iFogSim offers users a public constructor to create different types of Fog nodes. A sample code snippet to create heterogeneous Fog devices (nodes) on a particular hierarchical level is given below:

***Code Snippet-1***

- To be placed in Main Class

```
static int numOfFogDevices = 10;
static List<FogDevice> fogDevices = new ArrayList<FogDevice>();
static Map<String, Integer> getIdByName = new HashMap<String, Inte-
ger>();
private static void createFogDevices() {
    FogDevice cloud = createAFogDevice("cloud", 44800, 40000, 100,
10000, 0, 0.01, 16*103, 16*83.25);
    cloud.setParentId(-1);
    fogDevices.add(cloud);
    getIdByName.put(cloud.getName(), cloud.getId());
    for(int i=0;i<numOfFogDevices;i++){
        FogDevice device = createAFogDevice("FogDevice-"+i, get-
Value(12000, 15000), getValue(4000, 8000),
                   getValue(200, 300), getValue(500, 1000), 1, 0.01,
getValue(100,120), getValue(70, 75));
        device.setParentId(cloud.getId());
        device.setUplinkLatency(10);
        fogDevices.add(device);
        getIdByName.put(device.getName(), device.getId());}
}
private static FogDevice createAFogDevice(String nodeName, long mips,
int ram, long upBw, long downBw, int level, double ratePerMips, dou-
ble busyPower, double idlePower) {
    List<Pe> peList = new ArrayList<Pe>();
    peList.add(new Pe(0, new PeProvisionerOverbooking(mips)));
    int hostId = FogUtils.generateEntityId();
    long storage = 1000000;
    int bw = 10000;
    PowerHost host = new PowerHost(hostId, new RamProvisionerSim-
ple(ram), new BwProvisionerOverbooking(bw), storage, peList, new
```

```
StreamOperatorScheduler(peList), new FogLinearPowerModel(busyPower,
idlePower));
    List<Host> hostList = new ArrayList<Host>();
    hostList.add(host);
    String arch = "x86";
    String os = "Linux";
    String vmm = "Xen";
    double time_zone = 10.0;
    double cost = 3.0;
    double costPerMem = 0.05;
    double costPerStorage = 0.001;
    double costPerBw = 0.0;
    LinkedList<Storage> storageList = new LinkedList<Storage>();
    FogDeviceCharacteristics characteristics = new FogDeviceCharac-
teristics(arch, os, vmm, host, time_zone, cost,
                costPerMem, costPerStorage, costPerBw);
    FogDevice fogdevice = null;
    try {
        fogdevice = new FogDevice(nodeName, characteristics,
                    new AppModuleAllocationPolicy(hostList), stor-
ageList, 10, upBw, downBw, 0, ratePerMips);}
    catch (Exception e) {
        e.printStackTrace();}
    fogdevice.setLevel(level);
    return fogdevice;}
```

Code Snippet-1 creates a certain number of Fog nodes having configurations within a fixed range.

## *17.5.2 Create different application models.*

Different types of application model can be simulated through iFogSim. In following subsections, we discuss two types of such application model.

### 17.5.2.1 Master-Worker Application Model

The interaction among application modules on Master-Slave application model can be represented in Figure 17.3.

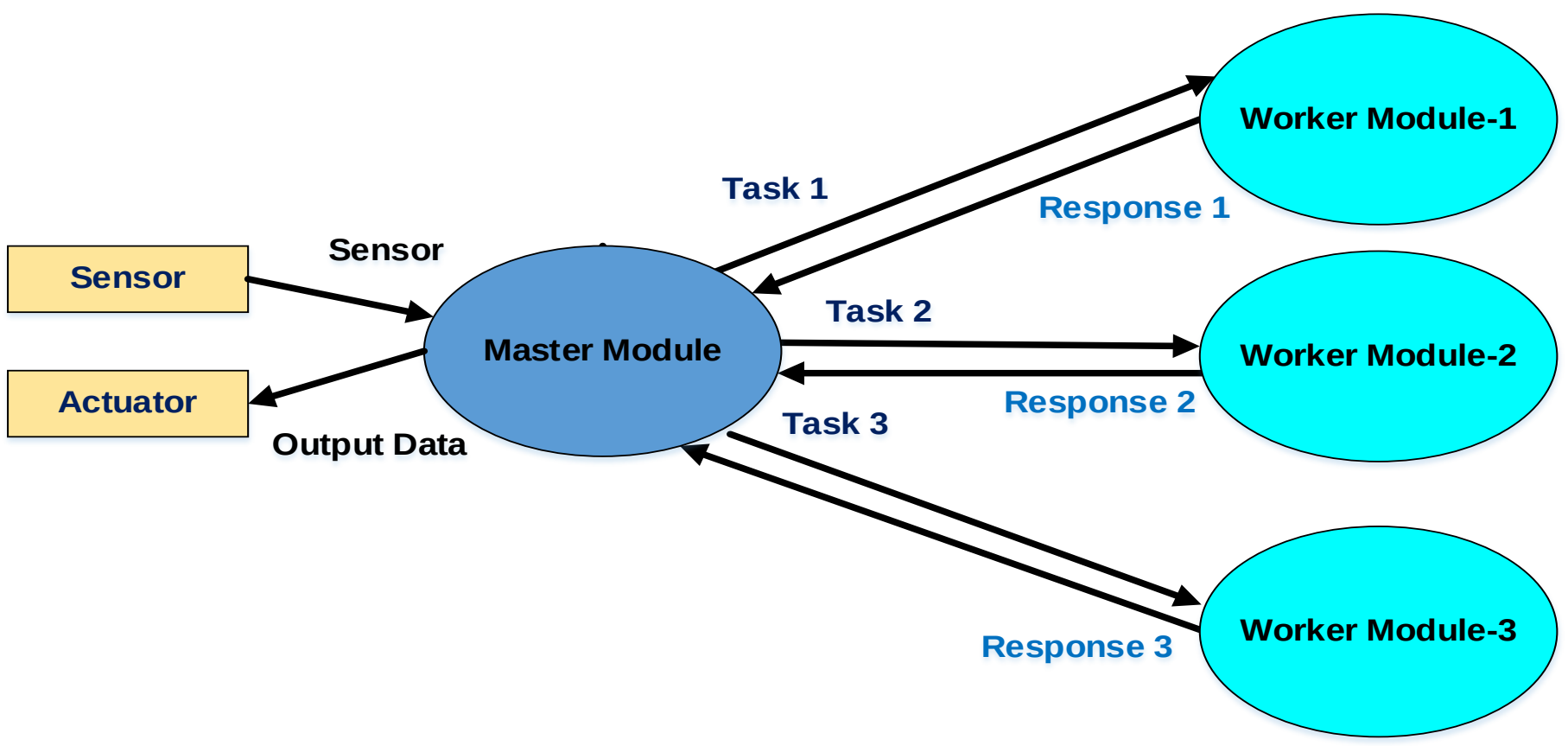

**Figure 17.3: Master-Worker application model**

To model such application in iFogSim, Code Snippet-2 can be used. Here, it is required to mention that, the name of an IoT sensor and the name of its emitted tuple type should be the same.

### *Code Snippet-2*

- To be placed in Main Class

```
private static Application createApplication(String appId, int bro-
kerId){
        Application application = Application.createApplication(ap-
pId, brokerId);
        application.addAppModule("MasterModule", 10);
        application.addAppModule("WorkerModule-1", 10);
        application.addAppModule("WorkerModule-2", 10);
        application.addAppModule("WorkerModule-3", 10);

        application.addAppEdge("Sensor", "MasterModule", 3000, 500,
"Sensor", Tuple.UP, AppEdge.SENSOR);
        application.addAppEdge("MasterModule", "WorkerModule-1", 100,
1000, "Task-1", Tuple.UP, AppEdge.MODULE);
        application.addAppEdge("MasterModule", "WorkerModule-2", 100,
1000, "Task-2", Tuple.UP, AppEdge.MODULE);
        application.addAppEdge("MasterModule", "WorkerModule-3", 100,
1000, "Task-3", Tuple.UP, AppEdge.MODULE);
        application.addAppEdge("WorkerModule-1", "MasterModule",20,
50, "Response-1", Tuple.DOWN, AppEdge.MODULE);
        application.addAppEdge("WorkerModule-2", "MasterModule",20,
50, "Response-2", Tuple.DOWN, AppEdge.MODULE);
        application.addAppEdge("WorkerModule-3", "MasterModule",20,
50, "Response-3", Tuple.DOWN, AppEdge.MODULE);
        application.addAppEdge("MasterModule", "Actuators", 100, 50,
"OutputData", Tuple.DOWN, AppEdge.ACTUATOR);

        application.addTupleMapping("MasterModule", " Sensor ",
"Task-1", new FractionalSelectivity(0.3));
        application.addTupleMapping("MasterModule", " Sensor ",
"Task-2", new FractionalSelectivity(0.3));
        application.addTupleMapping("MasterModule", " Sensor ",
"Task-3", new FractionalSelectivity(0.3));
        application.addTupleMapping("WorkerModule-1", "Task-1", "Re-
sponse-1", new FractionalSelectivity(1.0));
        application.addTupleMapping("WorkerModule-2", "Task-2", "Re-
sponse-2", new FractionalSelectivity(1.0));
        application.addTupleMapping("WorkerModule-3", "Task-3", "Re-
sponse-3", new FractionalSelectivity(1.0));
        application.addTupleMapping("MasterModule", "Response-1",
"OutputData", new FractionalSelectivity(0.3));
        application.addTupleMapping("MasterModule", "Response-2",
"OutputData", new FractionalSelectivity(0.3));
```

```
        application.addTupleMapping("MasterModule", "Response-3", 
"OutputData", new FractionalSelectivity(0.3));

        final AppLoop loop1 = new AppLoop(new Ar-
rayList<String>(){{add("Sensor");add("MasterModule");add("WorkerMod-
ule-1");add("MasterModule");add("Actuator");}});
        final AppLoop loop2 = new AppLoop(new Ar-
rayList<String>(){{add("Sensor");add("MasterModule");add("WorkerMod-
ule-2");add("MasterModule");add("Actuator");}});
        final AppLoop loop3 = new AppLoop(new Ar-
rayList<String>(){{add("Sensor");add("MasterModule");add("WorkerMod-
ule-3");add("MasterModule");add("Actuator");}});
        List<AppLoop> loops = new ArrayList<Ap-
pLoop>(){{add(loop1);add(loop2);add(loop3);}};
        application.setLoops(loops);

        return application;}
```

### 17.5.2.2 Sequential Unidirectional dataflow application model

Figure 17.4 depicts a sample sequential unidirectional application model.

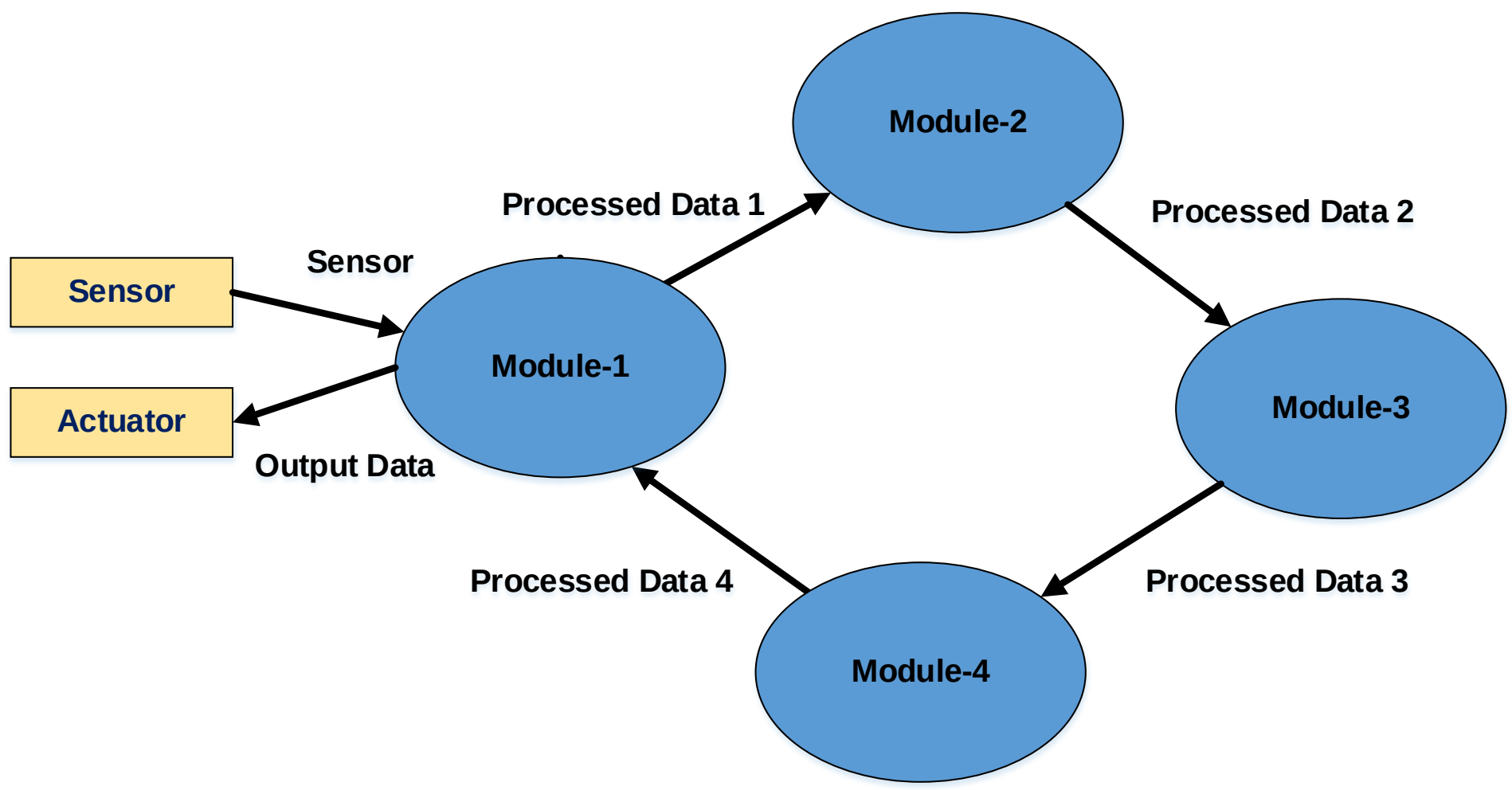

**Figure 17.4: Sequential Unidirectional dataflow application model**

Code Snippet-3 refers the instructions to model such applications in iFogSim.

***Code Snippet-3***

- To be placed in Main Class

```
private static Application createApplication(String appId, int bro-
kerId){
        Application application = Application.createApplication(ap-
pId, brokerId);
        application.addAppModule("Module1", 10);
        application.addAppModule("Module2", 10);
        application.addAppModule("Module3", 10);
        application.addAppModule("Module4", 10);
```

```
        application.addAppEdge("Sensor", "Module1", 3000, 500, "Sen-
sor", Tuple.UP, AppEdge.SENSOR);
        application.addAppEdge("Module1", "Module2", 100, 1000, "Pro-
cessedData-1", Tuple.UP, AppEdge.MODULE);
        application.addAppEdge("Module2", "Module3", 100, 1000, "Pro-
cessedData-2", Tuple.UP, AppEdge.MODULE);
        application.addAppEdge("Module3", "Module4", 100, 1000, "Pro-
cessedData-3", Tuple.UP, AppEdge.MODULE);
        application.addAppEdge("Module4", "Module1", 100, 1000, "Pro-
cessedData-4", Tuple.DOWN, AppEdge.MODULE);
        application.addAppEdge("Module1", "Actuators", 100, 50, "Out-
putData", Tuple.DOWN, AppEdge.ACTUATOR);

        application.addTupleMapping("Module1", "Sensor", "Processed-
Data-1", new FractionalSelectivity(1.0));
        application.addTupleMapping("Module2", "ProcessedData-1",
"ProcessedData-2", new FractionalSelectivity(1.0));
        application.addTupleMapping("Module3", "ProcessedData-2",
"ProcessedData-3", new FractionalSelectivity(1.0));
        application.addTupleMapping("Module4", "ProcessedData-3",
"ProcessedData-4", new FractionalSelectivity(1.0));
        application.addTupleMapping("Module1", "ProcessedData-4",
"OutputData", new FractionalSelectivity(1.0));

        final AppLoop loop1 = new AppLoop(new Ar-
rayList<String>(){{add("Sensor");add("Module1");add("Mod-
ule2");add("Module3");add("Module4");add("Module1");add("Actua-
tor");}});
        List<AppLoop> loops = new ArrayList<Ap-
pLoop>(){{add(loop1);}};
        application.setLoops(loops);
        return application;}
```

### *17.5.3 Application Modules with different configuration.*

The following Code Snippet-4 creates modules with different configurations.

***Code Snippet-4***

- To be placed in Main Class

```
private static Application createApplication(String appId, int bro-
kerId){
        Application application = Application.createApplication(ap-
pId, brokerId);
        application.addAppModule("ClientModule", 20,500, 1024, 1500);
        application.addAppModule("MainModule", 100, 1200, 4000, 100);

        application.addAppEdge("Sensor", "ClientModule", 3000, 500,
"Sensor", Tuple.UP, AppEdge.SENSOR);
        application.addAppEdge("ClientModule", "MainModule", 100,
1000, "PreProcessedData", Tuple.UP, AppEdge.MODULE);
        application.addAppEdge("MainModule", "ClientModule", 100,
1000, "ProcessedData", Tuple.DOWN, AppEdge.MODULE);
        application.addAppEdge("ClientModule", "Actuators", 100, 50,
"OutputData", Tuple.DOWN, AppEdge.ACTUATOR);
```

```
        application.addTupleMapping("ClientModule", "Sensor", "Pre-
ProcessedData", new FractionalSelectivity(1.0));
        application.addTupleMapping("MainModule", "PreProcessedData",
"ProcessedData", new FractionalSelectivity(1.0));
        application.addTupleMapping("ClientModule", "ProcessedData",
"OutputData", new FractionalSelectivity(1.0));

        final AppLoop loop1 = new AppLoop(new Ar-
rayList<String>(){{add("Sensor");add("ClientModule");add("MainMod-
ule");add("Actuator");}});
        List<AppLoop> loops = new ArrayList<Ap-
pLoop>(){{add(loop1);}};
        application.setLoops(loops);
        return application;}
```

- To be placed in Application Class

```
public void addAppModule(String moduleName, int ram, int mips, long
size, long bw){
        String vmm = "Xen";
        AppModule module = new AppModule(FogUtils.generateEntityId(),
moduleName, appId, userId,
                mips, ram, bw, size, vmm, new TupleScheduler(mips, 1),
new HashMap<Pair<String, String>, SelectivityModel>());
        getModules().add(module);
    }
```

## *17.5.4 Sensors with different tuple emission rate.*

To create sensors with different tuple emission rate, Code Snippet-5 can be used.

***Code Snippet-5***

- To be placed in Main Class

```
private static FogDevice addLowLevelFogDevice(String id, int bro-
kerId, String appId, int parentId){
        FogDevice lowLevelFogDevice = createAFogDevice("LowLevelFog-
Device-"+id, 1000, 1000, 10000, 270, 2, 0, 87.53, 82.44);
        lowLevelFogDevice.setParentId(parentId);
        getIdByName.put(lowLevelFogDevice.getName(), lowLevelFogDe-
    vice.getId());}
        Sensor sensor = new Sensor("s-"+id, "Sensor", brokerId, ap-
pId, new DeterministicDistribution(getValue(5.00)));
        sensors.add(sensor);
        Actuator actuator = new Actuator("a-"+id, brokerId, appId,
"OutputData");
        actuators.add(actuator);
        sensor.setGatewayDeviceId(lowLevelFogDevice.getId());
        sensor.setLatency(6.0);
        actuator.setGatewayDeviceId(lowLevelFogDevice.getId());
        actuator.setLatency(1.0);
        return lowLevelFogDevice;}

    private static double getValue(double min) {
        Random rn = new Random();
```

```
        return rn.nextDouble()*10 + min;}
```

### *17.5.5 Send specific number of tuples from a sensor.*

Code Snippet-6 enables sensors to create a specific number of tuples.

***Code Snippet-6***

- To be placed in Sensor Class

```
static int numOfMaxTuples = 100;
static int tuplesCount = 0;
public void transmit(){
    System.out.print(CloudSim.clock()+": ");
    if(tuplesCount<numOfMaxTuples){
        AppEdge _edge = null;
        for(AppEdge edge : getApp().getEdges()){
            if(edge.getSource().equals(getTupleType()))
                _edge = edge;
        }
        long cpuLength = (long) _edge.getTupleCpuLength();
        long nwLength = (long) _edge.getTupleNwLength();
        Tuple tuple = new Tuple(getAppId(), FogUtils.gener-
ateTupleId(), Tuple.UP, cpuLength, 1, nwLength, outputSize,
                    new UtilizationModelFull(), new UtilizationModel-
Full(), new UtilizationModelFull());
        tuple.setUserId(getUserId());
        tuple.setTupleType(getTupleType());
        tuple.setDestModuleName(_edge.getDestination());
        tuple.setSrcModuleName(getSensorName());
        Logger.debug(getName(), "Sending tuple with tupleId = "+tu-
ple.getCloudletId());
        int actualTupleId = updateTimings(getSensorName(), tu-
ple.getDestModuleName());
        tuple.setActualTupleId(actualTupleId);
        send(gatewayDeviceId, getLatency(), FogEvents.TUPLE_ARRI-
VAL,tuple);
        tuplesCount++;
    }
}
```

### *17.5.6 Mobility of a Fog device.*

In hierarchical order, each Fog device of particular level is connected with upper level Fog nodes. Code Snippet-7 represents how to deal with mobility issues in iFogSim. Here, we have considered mobility of arbitrary lower level Fog devices to a certain destination.

***Code Snippet-7***

- To be placed in Main Class

```
static Map<Integer, Pair<Double, Integer>> mobilityMap = new
HashMap<Integer, Pair<Double, Integer>>();
    static String mobilityDestination = "FogDevice-0";
    private static FogDevice addLowLevelFogDevice(String id, int bro-
kerId, String appId, int parentId){
        FogDevice lowLevelFogDevice = createAFogDevice("LowLevelFog-
Device-"+id, 1000, 1000, 10000, 270, 2, 0, 87.53, 82.44);
        lowLevelFogDevice.setParentId(parentId);
        getIdByName.put(lowLevelFogDevice.getName(), lowLevelFogDe-
vice.getId());

        if((int)(Math.random()*100)%2==0){
            Pair<Double, Integer> pair = new Pair<Double, Inte-
ger>(100.00, getIdByName.get(mobilityDestination));
            mobilityMap.put(lowLevelFogDevice.getId(), pair);}

        Sensor sensor = new Sensor("s-"+id, "Sensor", brokerId, ap-
pId, new DeterministicDistribution(getValue(5.00)));
        sensors.add(sensor);
        Actuator actuator = new Actuator("a-"+id, brokerId, appId,
"OutputData");
        actuators.add(actuator);
        sensor.setGatewayDeviceId(lowLevelFogDevice.getId());
        sensor.setLatency(6.0);
        actuator.setGatewayDeviceId(lowLevelFogDevice.getId());
        actuator.setLatency(1.0);
        return lowLevelFogDevice;}
```

- Inclusion in the Main method

```
Controller controller = new Controller("master-controller", fogDe-
vices, sensors, actuators);
controller.setMobilityMap(mobilityMap);
```

- To be placed in Controller Class

```
private static Map<Integer, Pair<Double, Integer>> mobilityMap;
public void setMobilityMap(Map<Integer, Pair<Double, Integer>> mobil-
ityMap) {
        this.mobilityMap = mobilityMap;
    }
    private void scheduleMobility(){
        for(int id: mobilityMap.keySet()){
            Pair<Double, Integer> pair = mobilityMap.get(id);
            double mobilityTime = pair.getFirst();
            int mobilityDestinationId = pair.getSecond();
            Pair<Integer, Integer> newConnection = new Pair<Integer,
Integer>(id, mobilityDestinationId);
            send(getId(), mobilityTime, FogEvents.FutureMobility,
newConnection);
        }
    }
```

```
private void manageMobility(SimEvent ev) {

        Pair<Integer, Integer>pair = (Pair<Integer, Integer>)ev.get-
Data();
        int deviceId = pair.getFirst();
        int newParentId = pair.getSecond();
        FogDevice deviceWithMobility = getFogDeviceById(deviceId);
        FogDevice mobilityDest = getFogDeviceById(newParentId);
        deviceWithMobility.setParentId(newParentId);
        System.out.println(CloudSim.clock()+" "+deviceWithMobil-
ity.getName()+" is now connected to "+mobilityDest.getName());}
```

- Inclusion in Controller startEntity method

```
scheduleMobility();
```

- Inclusion in Controller processEvent method

```
case FogEvents.FutureMobility:
            manageMobility(ev);
            break;
```

- To be placed in FogEvents Class

```
public static final int FutureMobility = BASE+26;
```

In Code Snippet-7, users can add other required instructions on manageMobility method to deal with the mobility driven issues such as AppModule migration and connection with latency.

### *17.5.7 Connect lower level Fog devices with nearby gateways*

Code Snippet-8 refers to a simple way to connect low level Fog devices to nearby gateway Fog devices. Here the gateway Fog devices are created with corresponding x and y coordinate values.

***Code Snippet-8***

- To be placed in Main Class

```
private static FogDevice addLowLevelFogDevice(String id, int bro-
kerId, String appId){
        FogDevice lowLevelFogDevice = createAFogDevice("LowLevelFog-
Device-"+id, 1000, 1000, 10000, 270, 2, 0, 87.53, 82.44);
        lowLevelFogDevice.setParentId(-1);
        lowLevelFogDevice.setxCoordinate(getValue(10.00));
        lowLevelFogDevice.setyCoordinate(getValue(15.00));
        getIdByName.put(lowLevelFogDevice.getName(), lowLevelFogDe-
vice.getId());
```

```
        Sensor sensor = new Sensor("s-"+id, "Sensor", brokerId, ap-
pId, new DeterministicDistribution(getValue(5.00)));
        sensors.add(sensor);
        Actuator actuator = new Actuator("a-"+id, brokerId, appId,
"OutputData");
        actuators.add(actuator);
        sensor.setGatewayDeviceId(lowLevelFogDevice.getId());
        sensor.setLatency(6.0);
        actuator.setGatewayDeviceId(lowLevelFogDevice.getId());
        actuator.setLatency(1.0);
        return lowLevelFogDevice;}

private static double getValue(double min) {
        Random rn = new Random();
        return rn.nextDouble()*10 + min;}
```

- To be placed in Constructor Class

```
private void gatewaySelection() {
        // TODO Auto-generated method stub
        for(int i=0;i<getFogDevices().size();i++){
            FogDevice fogDevice = getFogDevices().get(i);
            int parentID=-1;
            if(fogDevice.getParentId()==-1) {
                double minDistance = Config.MAX_NUMBER;
                for(int j=0;j<getFogDevices().size();j++){
                    FogDevice anUpperDevice = getFogDevices().get(j);
                    if(fogDevice.getLevel()+1==anUpperDe-
vice.getLevel()){
                        double distance = calculateDistance(fogDe-
vice,anUpperDevice);
                        if(distance<minDistance){
                            minDistance = distance;
                            parentID = anUpperDevice.getId();}
                    }
                }
            }
            fogDevice.setParentId(parentID);
        }
    }

private double calculateDistance(FogDevice fogDevice, FogDevice anUp-
perDevice) {
        // TODO Auto-generated method stub
        return Math.sqrt(Math.pow(fogDevice.getxCoordinate()-anUpper-
Device.getxCoordinate(), 2.00)+
                Math.pow(fogDevice.getyCoordinate()-anUpperDe-
vice.getyCoordinate(), 2.00));}
```

- To be placed in FogDevice Class

```
protected double xCoordinate;
protected double yCoordinate;

public double getxCoordinate() {
    return xCoordinate;}

public void setxCoordinate(double xCoordinate) {
```

```
    this.xCoordinate = xCoordinate;}

public double getyCoordinate() {
    return yCoordinate;}

public void setyCoordinate(double yCoordinate) {
    this.yCoordinate = yCoordinate;}
```

- Inclusion in Controller constructor method

```
gatewaySelection();
```

- Inclusion in Config Class

```
public static final double MAX_NUMBER = 9999999.00;
```

### *17.5.9 Make Cluster of Fog devices*

In Code Snippet-9, we draw a very simple principle for creating cluster of Fog devices. Here, two fog devices, residing in the same level and connected with identical upper level Fog nodes, if located at a threshold distance, they are considered belonging to the same Fog cluster.

***Code Snippet-9***

- To be placed in Controller Class

```
static Map<Integer, Integer> clusterInfo = new HashMap<Integer,Inte-
ger>();
static Map<Integer, List<Integer>> clusters = new HashMap<Integer,
List<Integer>>();
    private void formClusters() {
        for(FogDevice fd: getFogDevices()){
            clusterInfo.put(fd.getId(), -1);
        }

        int clusterId = 0;

        for(int i=0;i<getFogDevices().size();i++){
            FogDevice fd1 = getFogDevices().get(i);
            for(int j=0;j<getFogDevices().size();j++) {
                FogDevice fd2 = getFogDevices().get(j);
                if(fd1.getId()!=fd2.getId()&& fd1.getParen-
tId()==fd2.getParentId()
                       &&calculateDistance(fd1, fd2)<Config.CLUS-
TER_DISTANCE && fd1.getLevel()==fd2.getLevel())
                {
                    int fd1ClusteriD = clusterInfo.get(fd1.getId());
                    int fd2ClusteriD = clusterInfo.get(fd2.getId());
                    if(fd1ClusteriD==-1 && fd2ClusteriD==-1){
```

```
                        clusterId++;
                        clusterInfo.put(fd1.getId(), clusterId);
                        clusterInfo.put(fd2.getId(), clusterId);

                    }
                    else if(fd1ClusteriD==-1)
                        clusterInfo.put(fd1.getId(), cluster-
Info.get(fd2.getId()));
                    else if(fd2ClusteriD==-1)
                        clusterInfo.put(fd2.getId(), cluster-
Info.get(fd1.getId()));
                }
            }
        }

        for(int id:clusterInfo.keySet()){
            if(!clusters.containsKey(clusterInfo.get(id))){
                List<Integer>clusterMembers = new ArrayList<Inte-
ger>();
                clusterMembers.add(id);
                clusters.put(clusterInfo.get(id), clusterMembers);
            }
            else
            {
                List<Integer>clusterMembers = clusters.get(cluster-
Info.get(id));
                clusterMembers.add(id);
                clusters.put(clusterInfo.get(id), clusterMembers);
            }
        }

        for(int id:clusters.keySet())
            System.out.println(id+" "+clusters.get(id));

    }
```

- Inclusion in Controller constructor method:

```
formClusters();
```

- Inclusion in Config Class:

```
public static final double CLUSTER_DISTANCE = 2.00;
```

## 17.6 Simulation of a Placement Policy

In this section, we discuss a simple application placement scenario and implement the placement policy in iFogSim simulated Fog environment.

- ***Structure of physical environment:***

In the Fog environment, the devices are orchestrated in three-tire hierarchical order (Figure 17.5). The lower level End Fog devices are connected to the IoT sensors and actuators. The Gateway Fog devices brides the Cloud datacenter and end Fog devices to execute a modular application. For simplicity, the Fog devices of same hierarchical levels are considered homogeneous. The sensing frequency is same for all sensors.

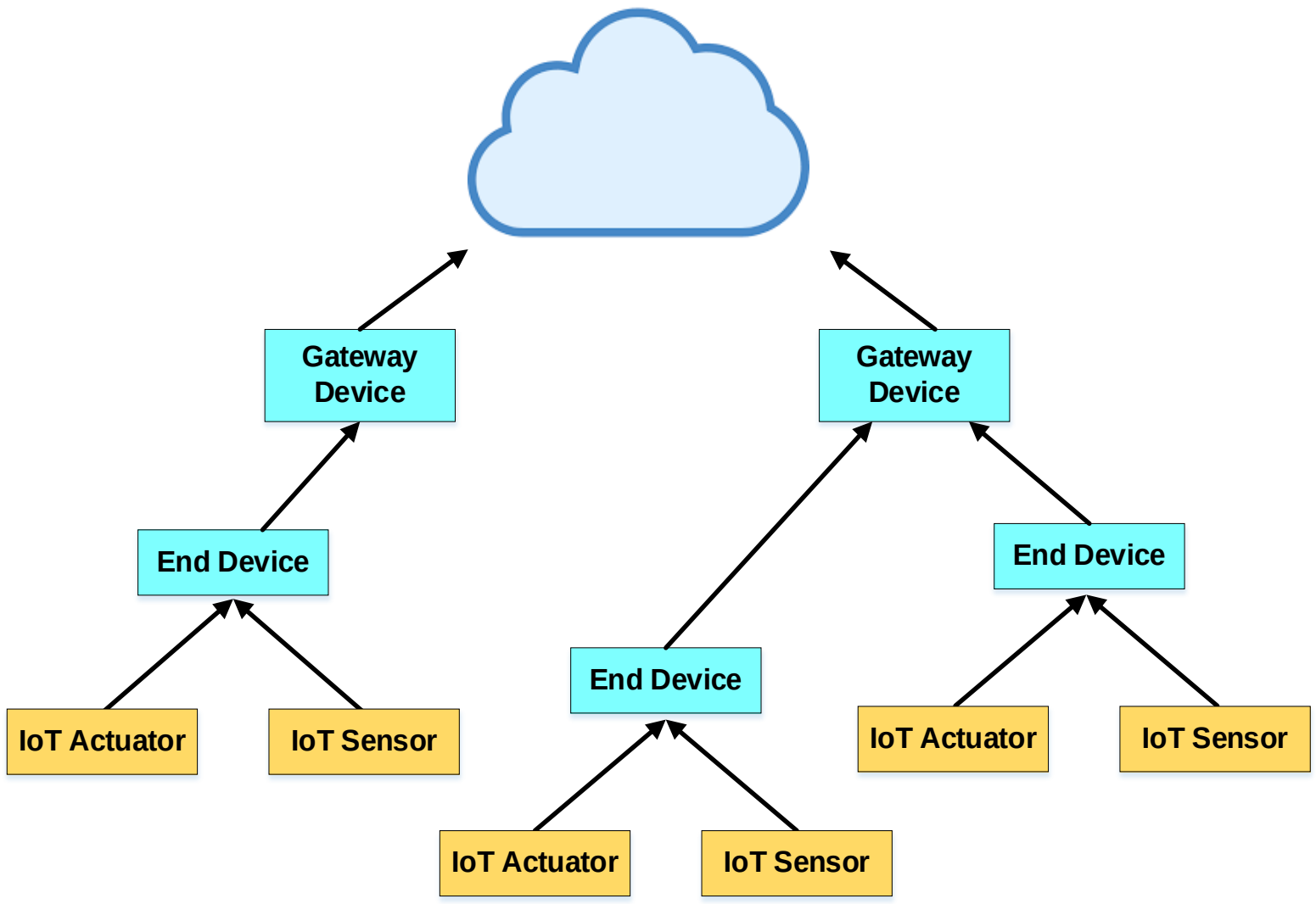

**Figure 17.5: Network topology for the placement policy**

- ***Assumptions for logical components:***

The application model is depicted in Figure 17.6. Here we assume that, ClientModule is placed in End Fog devices and StorageModule is placed in Cloud. The MainModule requires certain amount of computational resources to be initiated. To serve the demand of different End devices within their deadline, additional resources can be requested by End devices to connected Gateway Fog devices.

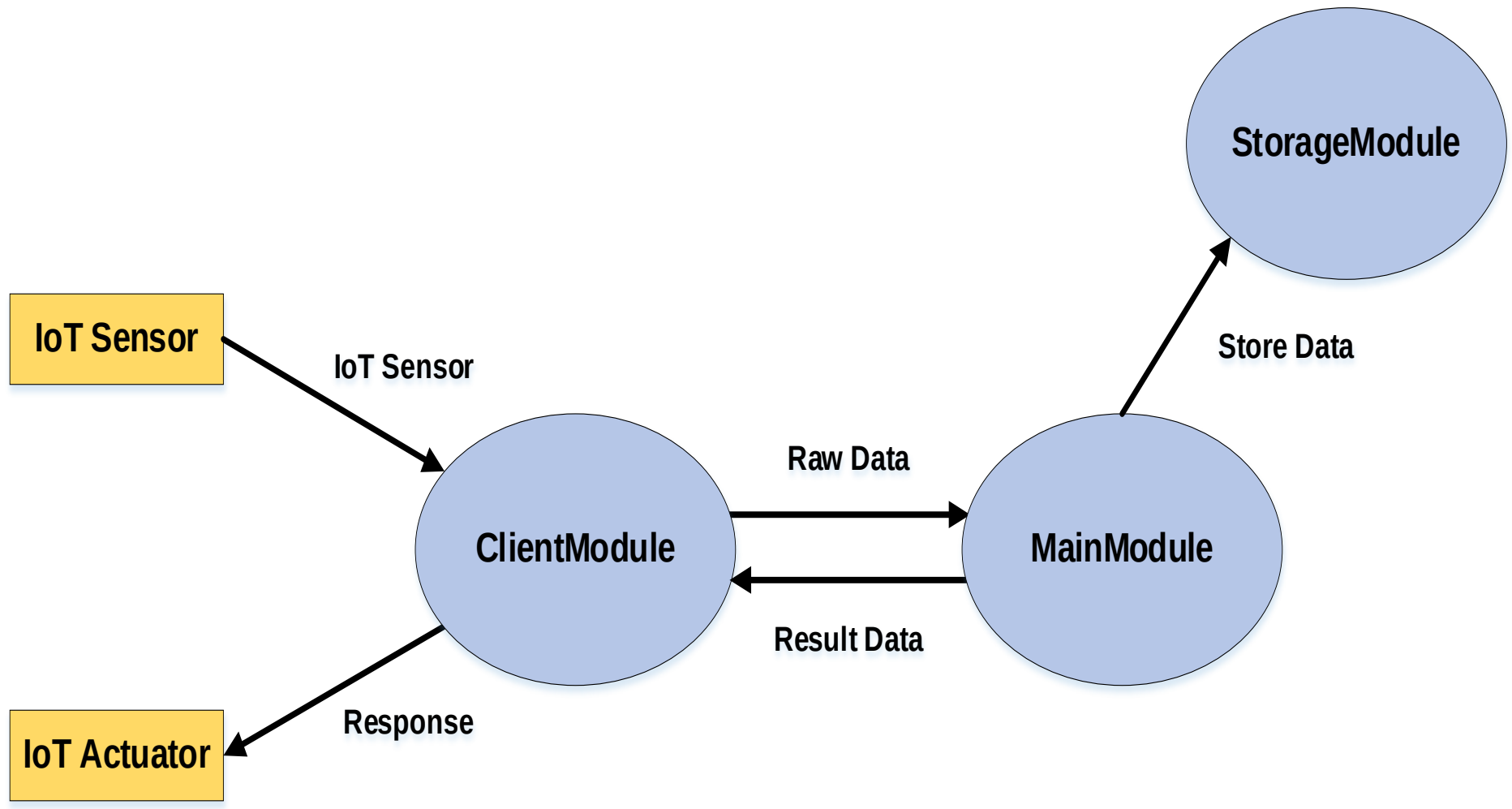

**Figure 17.6: Application model for the placement policy**

- ***Management (Application placement) policy:***

We target to place the MainApplication modules in Gateway Fog devices for different End devices based on their deadline requirement and resource availability in the host devices. For easier understanding, the flowchart of the application placement policy is represented in Figure 17.7.

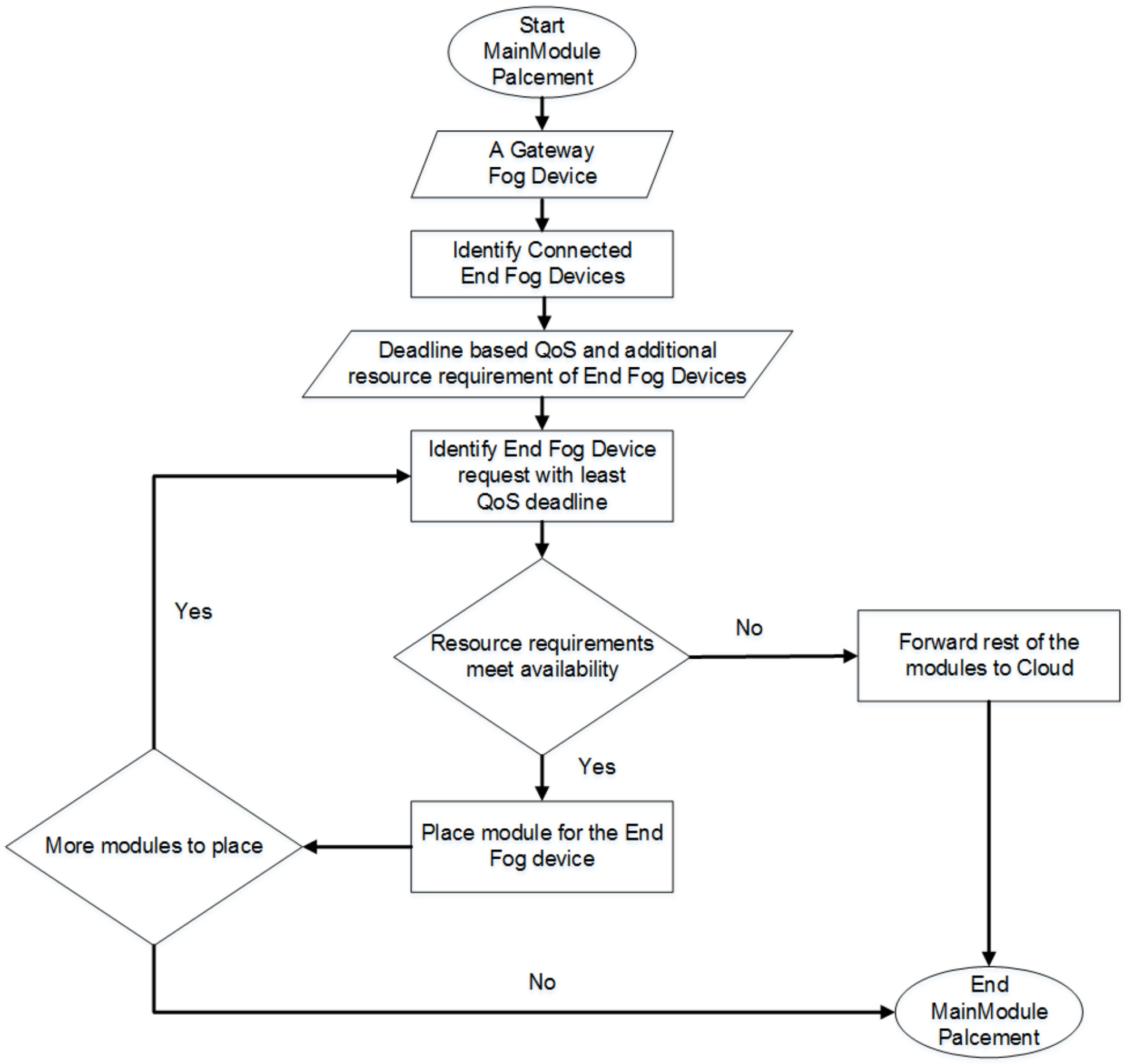

**Figure 17.7: Flow-chart of the application placement policy**

Code Snippet-10 represents necessary instruction to simulate the case scenario in iFog-Sim toolkit. Here MyApplication, MySensor, MyFogDevice, MyActuator, My Con-troller and MyPlacement class is same as Application, Sensor, FogDevice, Actuator, Contorller and ModulePacement class of iFogSim packages respectively. The inclu-sions are explicitly mentioned.

***Code Snippet-10***

- Main Class

```
public class TestApplication {
    static List<MyFogDevice> fogDevices = new ArrayList<MyFogDe-
vice>();
    static Map<Integer,MyFogDevice> deviceById = new HashMap<Inte-
ger,MyFogDevice>();
    static List<MySensor> sensors = new ArrayList<MySensor>();
    static List<MyActuator> actuators = new ArrayList<MyActuator>();
    static List<Integer> idOfEndDevices = new ArrayList<Integer>();
```

```
    static Map<Integer, Map<String, Double>> deadlineInfo = new
HashMap<Integer, Map<String, Double>>();
    static Map<Integer, Map<String, Integer>> additionalMipsInfo =
new HashMap<Integer, Map<String, Integer>>();

    static boolean CLOUD = false;

    static int numOfGateways = 2;
    static int numOfEndDevPerGateway = 3;
    static double sensingInterval = 5;

    public static void main(String[] args) {

        Log.printLine("Starting TestApplication...");

        try {
            Log.disable();
            int num_user = 1;
            Calendar calendar = Calendar.getInstance();
            boolean trace_flag = false;
            CloudSim.init(num_user, calendar, trace_flag);
            String appId = "test_app";
            FogBroker broker = new FogBroker("broker");

            createFogDevices(broker.getId(), appId);

            MyApplication application = createApplication(appId, bro-
ker.getId());
            application.setUserId(broker.getId());

            ModuleMapping moduleMapping = ModuleMapping.createModule-
Mapping();

            moduleMapping.addModuleToDevice("storageModule",
"cloud");
            for(int i=0;i<idOfEndDevices.size();i++)
            {
                MyFogDevice fogDevice = deviceById.get(idOfEndDe-
vices.get(i));
                moduleMapping.addModuleToDevice("clientModule", fogDe-
vice.getName());
            }


            MyController controller = new MyController("master-con-
troller", fogDevices, sensors, actuators);

            controller.submitApplication(application, 0, new MyModu-
lePlacement(fogDevices, sensors, actuators, application, module-
Mapping,"mainModule"));

            TimeKeeper.getInstance().setSimulationStartTime(Calen-
dar.getInstance().getTimeInMillis());

            CloudSim.startSimulation();

            CloudSim.stopSimulation();

            Log.printLine("TestApplication finished!");
```

```
        } catch (Exception e) {
            e.printStackTrace();
            Log.printLine("Unwanted errors happen");
        }
    }

    private static double getvalue(double min, double max)
    {
        Random r = new Random();
        double randomValue = min + (max - min) * r.nextDouble();
        return randomValue;
    }

    private static int getvalue(int min, int max)
    {
        Random r = new Random();
        int randomValue = min + r.nextInt()%(max - min);
        return randomValue;
    }

    private static void createFogDevices(int userId, String appId) {
        MyFogDevice cloud = createFogDevice("cloud", 44800, 40000,
100, 10000, 0, 0.01, 16*103, 16*83.25);
        cloud.setParentId(-1);
        fogDevices.add(cloud);
        deviceById.put(cloud.getId(), cloud);

        for(int i=0;i<numOfGateways;i++){
            addGw(i+"", userId, appId, cloud.getId());
        }
    }

    private static void addGw(String gwPartialName, int userId,
String appId, int parentId){
        MyFogDevice gw = createFogDevice("g-"+gwPartialName, 2800,
4000, 10000, 10000, 1, 0.0, 107.339, 83.4333);
        fogDevices.add(gw);
        deviceById.put(gw.getId(), gw);
        gw.setParentId(parentId);
        gw.setUplinkLatency(4);
        for(int i=0;i<numOfEndDevPerGateway;i++){
            String endPartialName = gwPartialName+"-"+i;
            MyFogDevice end  = addEnd(endPartialName, userId, appId,
gw.getId());
            end.setUplinkLatency(2);
            fogDevices.add(end);
            deviceById.put(end.getId(), end);
        }
    }

    private static MyFogDevice addEnd(String endPartialName, int
userId, String appId, int parentId){
        MyFogDevice end = createFogDevice("e-"+endPartialName, 3200,
1000, 10000, 270, 2, 0, 87.53, 82.44);
        end.setParentId(parentId);
        idOfEndDevices.add(end.getId());
        MySensor sensor = new MySensor("s-"+endPartialName, "IoTSen-
sor", userId, appId, new DeterministicDistribution(sensingInterval));
```

```
// inter-transmission time of EEG sensor follows a deterministic dis-
tribution
        sensors.add(sensor);
        MyActuator actuator = new MyActuator("a-"+endPartialName,
userId, appId, "IoTActuator");
        actuators.add(actuator);
        sensor.setGatewayDeviceId(end.getId());
        sensor.setLatency(6.0);  // latency of connection between EEG
sensors and the parent Smartphone is 6 ms
        actuator.setGatewayDeviceId(end.getId());
        actuator.setLatency(1.0);  // latency of connection between
Display actuator and the parent Smartphone is 1 ms
        return end;
    }

    private static MyFogDevice createFogDevice(String nodeName, long
mips,
            int ram, long upBw, long downBw, int level, double rate-
PerMips, double busyPower, double idlePower) {
        List<Pe> peList = new ArrayList<Pe>();
        peList.add(new Pe(0, new PeProvisionerOverbooking(mips)));
        int hostId = FogUtils.generateEntityId();
        long storage = 1000000;
        int bw = 10000;

        PowerHost host = new PowerHost(
                hostId,
                new RamProvisionerSimple(ram),
                new BwProvisionerOverbooking(bw),
                storage,
                peList,
                new StreamOperatorScheduler(peList),
                new FogLinearPowerModel(busyPower, idlePower)
            );
        List<Host> hostList = new ArrayList<Host>();
        hostList.add(host);
        String arch = "x86";
        String os = "Linux";
        String vmm = "Xen";
        double time_zone = 10.0;
        double cost = 3.0;
        double costPerMem = 0.05;
        double costPerStorage = 0.001;
        double costPerBw = 0.0;
        LinkedList<Storage> storageList = new LinkedList<Storage>();
        FogDeviceCharacteristics characteristics = new FogDeviceChar-
acteristics(
                arch, os, vmm, host, time_zone, cost, costPerMem,
                costPerStorage, costPerBw);

        MyFogDevice fogdevice = null;
        try {
            fogdevice = new MyFogDevice(nodeName, characteristics,
                    new AppModuleAllocationPolicy(hostList), stor-
ageList, 10, upBw, downBw, 0, ratePerMips);
        } catch (Exception e) {
            e.printStackTrace();}
        fogdevice.setLevel(level);
        fogdevice.setMips((int) mips);
```

```
        return fogdevice;}

    @SuppressWarnings({"serial" })
    private static MyApplication createApplication(String appId, int 
userId){

        MyApplication application = MyApplication.createApplica-
tion(appId, userId);
        application.addAppModule("clientModule",10, 1000, 1000, 100);
        application.addAppModule("mainModule", 50, 1500, 4000, 800);
        application.addAppModule("storageModule", 10, 50, 12000, 
100);

        application.addAppEdge("IoTSensor", "clientModule", 100, 200, 
"IoTSensor", Tuple.UP, AppEdge.SENSOR);
        application.addAppEdge("clientModule", "mainModule", 6000, 
600  , "RawData", Tuple.UP, AppEdge.MODULE);
        application.addAppEdge("mainModule", "storageModule", 1000, 
300, "StoreData", Tuple.UP, AppEdge.MODULE);
        application.addAppEdge("mainModule", "clientModule", 100, 50, 
"ResultData", Tuple.DOWN, AppEdge.MODULE);
        application.addAppEdge("clientModule", "IoTActuator", 100, 
50, "Response", Tuple.DOWN, AppEdge.ACTUATOR);

        application.addTupleMapping("clientModule", "IoTSensor", 
"RawData", new FractionalSelectivity(1.0));
        application.addTupleMapping("mainModule", "RawData", "Result-
Data", new FractionalSelectivity(1.0));
        application.addTupleMapping("mainModule", "RawData", 
"StoreData", new FractionalSelectivity(1.0));
        application.addTupleMapping("clientModule", "ResultData", 
"Response", new FractionalSelectivity(1.0));

        for(int id:idOfEndDevices)
        {
            Map<String,Double>moduleDeadline = new 
HashMap<String,Double>();
            moduleDeadline.put("mainModule", getvalue(3.00, 5.00));
            Map<String,Integer>moduleAddMips = new HashMap<String,In-
teger>();
            moduleAddMips.put("mainModule", getvalue(0, 500));
            deadlineInfo.put(id, moduleDeadline);
            additionalMipsInfo.put(id,moduleAddMips);}

        final AppLoop loop1 = new AppLoop(new Ar-
rayList<String>(){{add("IoTSensor");add("clientModule");add("mainMod-
ule");add("clientModule");add("IoTActuator");}});
        List<AppLoop> loops = new ArrayList<Ap-
pLoop>(){{add(loop1);}};
        application.setLoops(loops);
        application.setDeadlineInfo(deadlineInfo);
        application.setAdditionalMipsInfo(additionalMipsInfo);
        return application;}
}
```

- Inclusion in MyApplication Class

```
private Map<Integer, Map<String, Double>> deadlineInfo;
private Map<Integer, Map<String, Integer>> additionalMipsInfo;
```

```
public Map<Integer, Map<String, Integer>> getAdditionalMipsInfo() {
        return additionalMipsInfo;
    }
public void setAdditionalMipsInfo(
        Map<Integer, Map<String, Integer>> additionalMipsInfo) {
        this.additionalMipsInfo = additionalMipsInfo;
    }
public void setDeadlineInfo(Map<Integer, Map<String, Double>> dead-
lineInfo) {
        this.deadlineInfo = deadlineInfo;
    }

public Map<Integer, Map<String, Double>> getDeadlineInfo() {
        return deadlineInfo;
    }
public void addAppModule(String moduleName,int ram, int mips, long
size, long bw){
        String vmm = "Xen";
        AppModule module = new AppModule(FogUtils.generateEntityId(),
moduleName, appId, userId,
                mips, ram, bw, size, vmm, new TupleScheduler(mips, 1),
new HashMap<Pair<String, String>, SelectivityModel>());

        getModules().add(module);  }
```

- Inclusion in MyFogDevice Class

```
private int mips;

    public int getMips() {
        return mips;
    }

    public void setMips(int mips) {
        this.mips = mips;
    }
```

- MyModulePlacement Class

```
public class MyModulePlacement extends MyPlacement{

protected ModuleMapping moduleMapping;
protected List<MySensor> sensors;
protected List<MyActuator> actuators;
protected String moduleToPlace;
protected Map<Integer, Integer> deviceMipsInfo;

public MyModulePlacement(List<MyFogDevice> fogDevices, List<MySensor>
sensors, List<MyActuator> actuators,
            MyApplication application, ModuleMapping moduleMapping,
String moduleToPlace){
        this.setMyFogDevices(fogDevices);
        this.setMyApplication(application);
        this.setModuleMapping(moduleMapping);
        this.setModuleToDeviceMap(new HashMap<String, List<Inte-
ger>>());
```

```
        this.setDeviceToModuleMap(new HashMap<Integer, List<AppMod-
ule>>());
        setMySensors(sensors);
        setMyActuators(actuators);
        this.moduleToPlace = moduleToPlace;
        this.deviceMipsInfo = new HashMap<Integer, Integer>();
        mapModules();
    }

    @Override
    protected void mapModules() {

        for(String deviceName : getModuleMapping().getModule-
Mapping().keySet()){
            for(String moduleName : getModuleMapping().getModule-
Mapping().get(deviceName)){
                int deviceId = CloudSim.getEntityId(deviceName);
                AppModule appModule = getMyApplication().getModuleBy-
Name(moduleName);
                if(!getDeviceToModuleMap().containsKey(deviceId))
                {
                    List<AppModule>placedModules = new Ar-
rayList<AppModule>();
                    placedModules.add(appModule);
                    getDeviceToModuleMap().put(deviceId, placedMod-
ules);
                }
                else
                {
                    List<AppModule>placedModules = getDeviceTo-
ModuleMap().get(deviceId);
                    placedModules.add(appModule);
                    getDeviceToModuleMap().put(deviceId, placedMod-
ules);
                }
            }
        }
        for(MyFogDevice device:getMyFogDevices())
        {
            int deviceParent = -1;
            List<Integer>children = new ArrayList<Integer>();

            if(device.getLevel()==1)
            {
                if(!deviceMipsInfo.containsKey(device.getId()))
                    deviceMipsInfo.put(device.getId(), 0);
                deviceParent = device.getParentId();
                for(MyFogDevice deviceChild:getMyFogDevices())
                {
                    if(deviceChild.getParentId()==device.getId()){
                        children.add(deviceChild.getId());}
                }
                Map<Integer, Double>childDeadline = new HashMap<Inte-
ger, Double>();
                for(int childId:children)
                    childDeadline.put(childId,getMyApplica-
tion().getDeadlineInfo().get(childId).get(moduleToPlace));
```

```
                List<Integer> keys = new ArrayList<Integer>(childDead-
line.keySet());

                for(int i = 0; i<keys.size()-1; i++)
                {
                    for(int j=0;j<keys.size()-i-1;j++)
                    {
                        if(childDeadline.get(keys.get(j))>childDead-
line.get(keys.get(j+1))){
                            int tempJ = keys.get(j);
                            int tempJn = keys.get(j+1);
                            keys.set(j, tempJn);
                            keys.set(j+1, tempJ);
                        }
                    }
                }
                int baseMipsOfPlacingModule = (int)getMyApplica-
tion().getModuleByName(moduleToPlace).getMips();
                for(int key:keys)
                {
                    int currentMips = deviceMipsInfo.get(de-
vice.getId());
                    AppModule appModule = getMyApplication().getMod-
uleByName(moduleToPlace);
                    int additionalMips = getMyApplication().getAddi-
tionalMipsInfo().get(key).get(moduleToPlace);
                    if(currentMips+baseMipsOfPlacingModule+addition-
alMips<device.getMips())
                    {
                        currentMips = currentMips+baseMipsOfPlacing-
Module+additionalMips;
                        deviceMipsInfo.put(device.getId(), current-
Mips);
                        if(!getDeviceToModuleMap().containsKey(de-
vice.getId()))
                        {
                            List<AppModule>placedModules = new Ar-
rayList<AppModule>();
                            placedModules.add(appModule);
                            getDeviceToModuleMap().put(device.getId(),
placedModules);

                        }
                        else
                        {
                            List<AppModule>placedModules = getDe-
viceToModuleMap().get(device.getId());
                            placedModules.add(appModule);
                            getDeviceToModuleMap().put(device.getId(),
placedModules);
                        }

                    }
                    else
                    {
                        List<AppModule>placedModules = getDeviceTo-
ModuleMap().get(deviceParent);
                        placedModules.add(appModule);
```

```
                        getDeviceToModuleMap().put(deviceParent,
placedModules);
                    }
                }
            }
        }
    }

    public ModuleMapping getModuleMapping() {
        return moduleMapping;
    }

    public void setModuleMapping(ModuleMapping moduleMapping) {
        this.moduleMapping = moduleMapping;
    }


    public List<MySensor> getMySensors() {
        return sensors;
    }

    public void setMySensors(List<MySensor> sensors) {
        this.sensors = sensors;
    }

    public List<MyActuator> getMyActuators() {
        return actuators;
    }

    public void setMyActuators(List<MyActuator> actuators) {
        this.actuators = actuators;
    }
}
```

## 17.7 A Case Study in Smart Healthcare

In Healthcare solution, hand held or body connected IoT devices for example; pulse oximeter, ECG monitor, smart watches, etc. perceive health context of the users through a Client application module. The IoT devices are usually connected with smart phones. The smart phones act as the Application gateway node for the corresponding application. These nodes pre-process the IoT-device sensed data. If resource availability in Application gateway node meets the requirements, the data analysis and event management operation of the application is conducted there otherwise the operations are executed in upper level Fog computational nodes. For the second case, Application gateway nodes select suitable computational nodes to deploy other application modules and initiates actuators based on the result coming from those modules. Extending such case of IoT-enabled healthcare solution [5], we discuss the way to simulate the corresponding Fog environment in iFogSim. The system architecture and application model for the IoT enabled healthcare solutions is represented in Figure 17.8 and 17.9 respectively. Features of the system and the application along with required guidelines to model them in iFogSim are listed below:

- It is an n-tire hierarchical Fog environment. As the rank of Fog levels goes higher, the number of Fog devices residing that level gets lower. Fog devices form Cluster among themselves and can be mobile. IoT devices (pulse oximeter, ECG monitor, etc.) are connected to lower level Fog devices. The sensing frequency of the IoT devices are different. Steps to model these physical entities are:

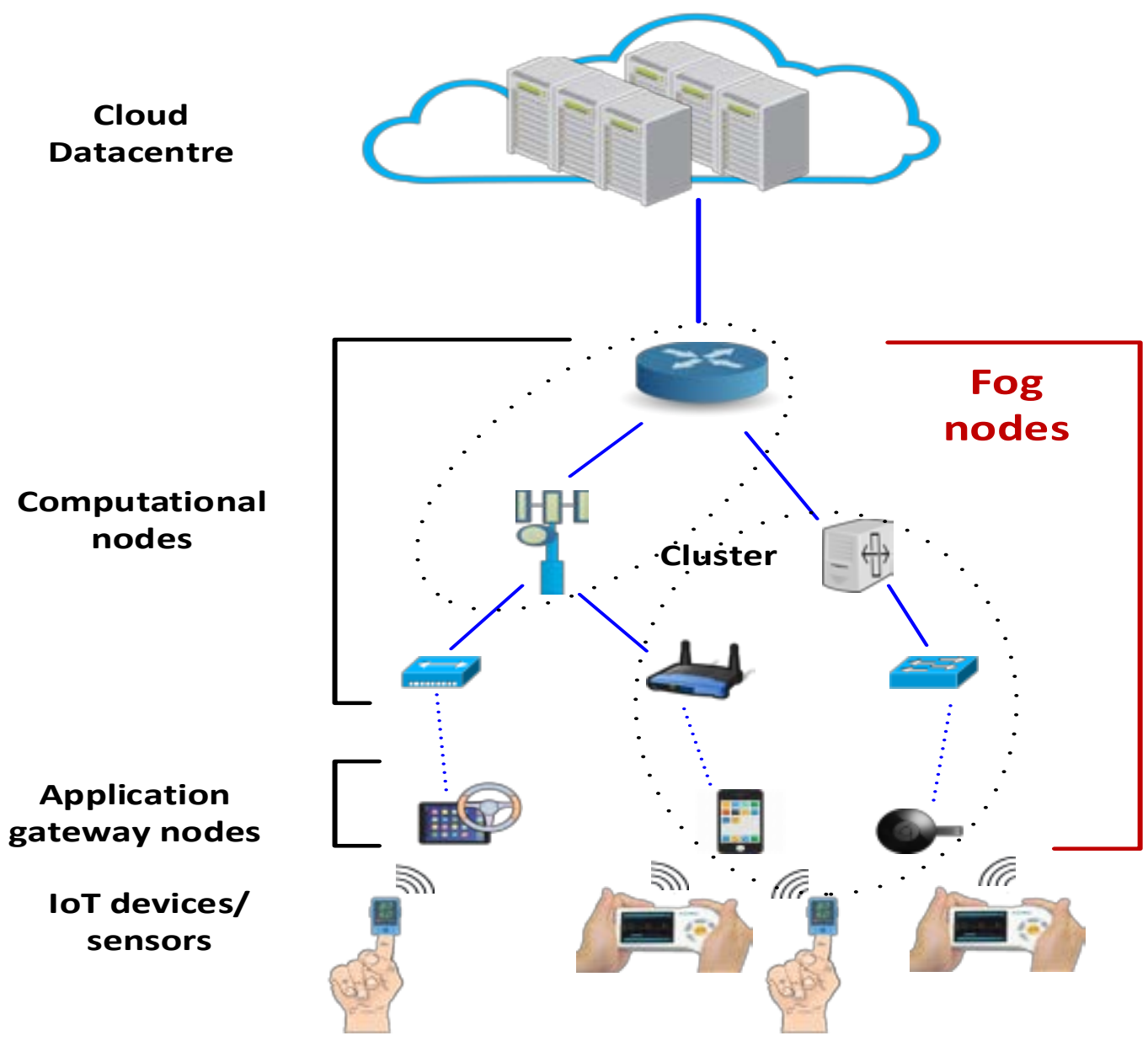

**Figure 17.8: Fog environment for IoT-enabled healthcare case study.**

1. Create FogDevice object and define n-tire hierarchical Fog environment by following Code Snippet 1 and 10.
2. Create Sensor object with different sensing interval and transmission of a particular number of Tuples using Code Snippet 5 and 6.

3. Model mobility of and form cluster of the Fog devices by modifying Code Snippet 7 and 9 respectively.

- The application model consists of 4 modules with a sequential unidirectional data flow. The requirements of the application modules are different and each application modules can request for additional resources from the host Fog devices to process a data within QoS-defined deadline. Steps to model these logical entities are:

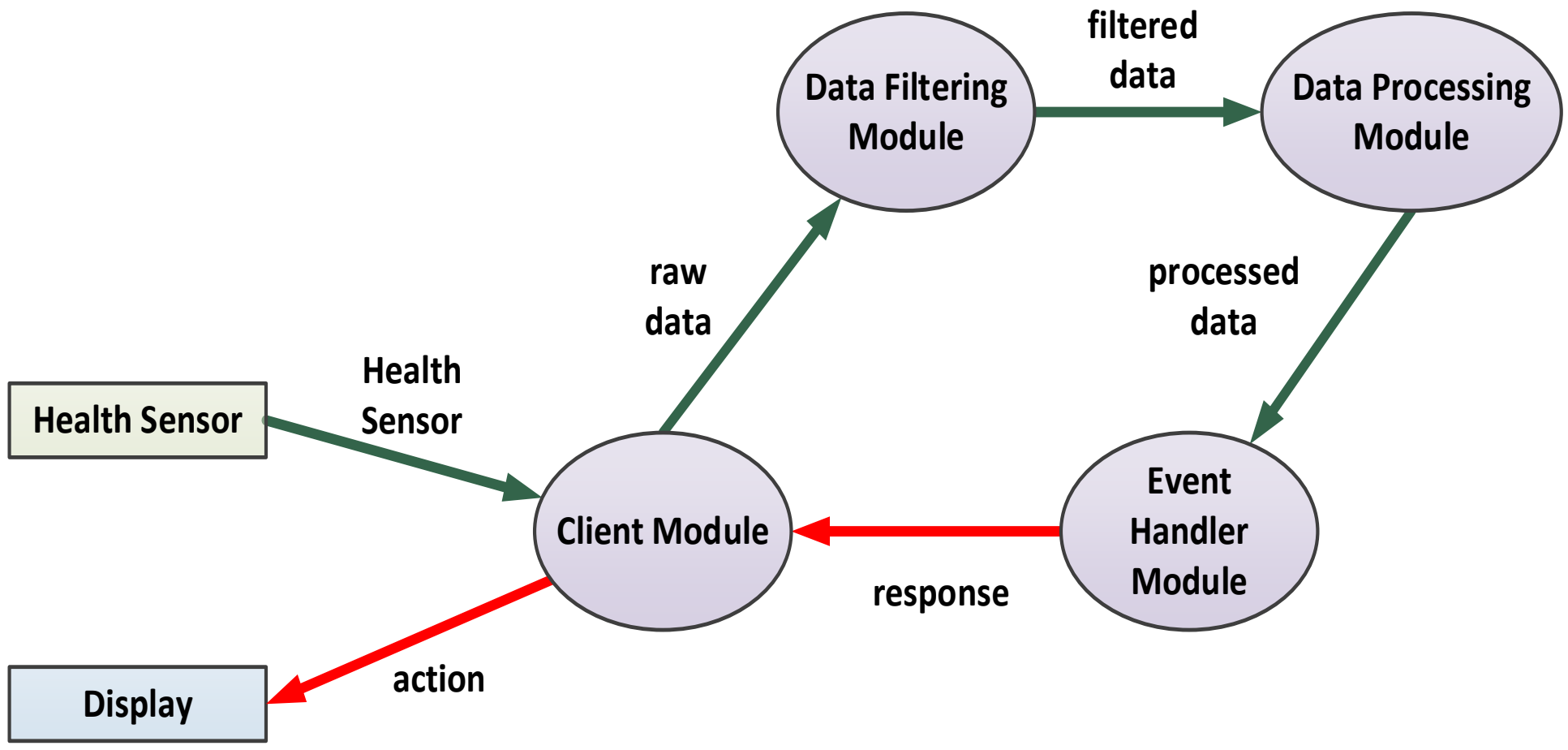

**Figure 17.9: Application Model for IoT-enabled healthcare case study.**

1. Define Application object for the discussing IoT-enabled healthcare application through Code Snippet 2 and 3.
2. Create ApplicationModule object with different requirements using Code Snippet 4.
3. Deal with additional requirements and deadline expectations of the ApplicationModule objects following Code Snippet 10.

- The application module placement in this case study should be done is such a way that takes least possible amount of time for the application to generate response for an event. In this case latency-aware placement of the modules on constrained Fog devices can be very effective [14]. Steps to model these management issues are:

1. Connect Application gateway nodes with low latency Fog computational nodes modifying Code Snippet 8.
2. Implement user defined latency-aware application module placement policy following Code Snippet 10.

## 17.8 Summary

In this chapter, we highlighted key features of iFogSim along with providing instructions to install it and simulate a Fog environment. We discussed some example scenarios and corresponding code snippets. Finally, demonstrated how to implement custom application placement in iFogSim simulated Fog environment along with a IoT-enabled smart healthcare case study.

The simulation source codes of example scenarios and placement policy discussed in this chapter are available from CLOUDS Laboratory GitHub webpage https://github.com/Cloudslab/iFogSimTutorials